# Uncovering Students' Mental Models of Generative Artificial Intelligence

Amrita Ganguly
*Information Sciences and Technology*
*George Mason University*
Fairfax, VA, USA
agangul@gmu.edu

Sai Sharanya Garika
*Information Sciences and Technology*
*George Mason University*
Fairfax, VA, USA
sgarika@gmu.edu

Aditya Johri
*Information Sciences and Technology*
*George Mason University*
Fairfax, VA, USA
johri@gmu.edu

***Abstract*—In this paper we present a study of students' mental models of generative AI (GenAI). A student's mental model of GenAI influences not only how they perceive the technology's capabilities and limitations but also how they choose to integrate it into their academic work. Whether they view it as a collaborative partner, a shortcut to complete tasks, or something in between, depends on how they conceptualize its use. This study addresses the following questions: (I) What mental models do undergraduate students hold about GenAI? and (II) What aspects of conceptual knowledge - declarative, procedural, and conditional - are present in these mental models? Sixty-four concept maps were collected from students enrolled in a course on technology ethics. Students were asked to construct concept maps representing their understanding of GenAI use. The concept maps were analyzed using a structured codebook and the analysis revealed five categories of mental models: technical process based, educational tool based, transition model, consequence aware model, integrated model. Declarative knowledge was most dominant across maps, suggesting that students largely understood GenAI primarily at a surface level - knowing its names, tools, and applications but demonstrate limited procedural understanding of how it works and limited conditional knowledge about when and why it should or should not be used. By identifying students' mental models, we can improve students' AI literacy by designing curriculum and guidelines that improve cognition while ensuring responsible and ethical use.**

## I. Introduction

Mental models are internal cognitive frameworks used to reason and make decisions and can be the basis of individual behaviors with a system. They are constructed by individuals based on their unique life experiences, perceptions, and understandings of the world [1]. In the context of educational technology use, mental models are significant because they fundamentally shape how learners understand new concepts, make decisions about technology use, and develop behavioral patterns around emerging tools. According to Carroll & Olson [2] mental models indicate utility over accuracy and effectiveness.

GenAI applications such as ChatGPT, Claude, and GitHub Copilot are rapidly being integrated into higher education, adding a new dimension to how students think about and use AI. Students actively engage with GenAI for academic purposes, including improving assignment drafts, generating ideas, and completing tasks. Such use can improve short-term task performance, for example, by increasing essay scores [3,4]. Discrepancies between students' mental models and their use though can lead to significant problems including misunderstandings about appropriate use, unintentional policy violations, missed opportunities for productive engagement with technology, or resistance to institutional directives that seem disconnected from students' understanding of GenAI.

Prior research on GenAI adoption in education has focused on quantitative assessments of knowledge, measuring what students know about AI concepts through tests and surveys, or examining patterns of tool adoption and technical skill development. While valuable, these approaches provide an incomplete picture because they rarely investigate the underlying cognitive structures, reasoning processes, and intuitive frameworks that students use to make sense of GenAI. Exploring these student perspectives is essential for designing pedagogical approaches and institutional policies that resonate with students' actual thinking patterns, address their misconceptions, and build on their existing conceptual foundations.

According to Gilbert and Boulter [5], concept maps serve as vital "expressed models" that allow researchers to externalize and investigate an individual's internal, dynamic conceptual framework or mental model. These visual tools provide a powerful means to probe complex and abstract concepts - such as technical mechanisms or ethical values - that are often difficult to verbalize, effectively making implicit beliefs and prior knowledge cognitively accessible [6]. Analysis of these maps through various approaches helps identify the degree of knowledge integration and specific gaps in a student's conceptual systems [7]. Furthermore, incorporating drawing activities within the mapping process engages distinct cognitive processes, enabling participants to reveal subconscious values, beliefs, and misconceptions that might remain hidden in traditional interview settings [8].

In this WIP research study, we investigated undergraduate students' mental models of GenAI by leveraging concept mapping as a primary elicitation method. These maps not only surface the core conceptual components students associate with GenAI, but also illuminate the connections, hierarchies, and reasoning patterns that reflect their understanding. Prior research has employed various approaches to elicit mental models such as think-aloud, storyboarding, and structured interviews [9, 10, 11]. Therefore, we position this analysis as an initial step in mental model elicitation, providing an overview that can be further validated through complementary methods such as interviews or think-aloud studies. This study contributes to engineering education by using concept maps to capture students' mental models of GenAI, providing insights into how undergraduates understand GenAI, and identifying knowledge gaps that can inform instructional design. Guided by these contributions, we address two research questions: (1) What mental models of GenAI are reflected in undergraduate students' concept maps, and (2) What types of knowledge – declarative, procedural, and conditional – shape these mental models?

## II. Methodology

### A. Data collection

Data was collected from 86 undergraduate students in a required technology ethics course within the IT curriculum As part of an assignment. Students were instructed to create concept maps to represent their understanding of GenAI and to write a brief explanations. No other specific instructions given to avoid constraining their thinking. As technology students, they already had a basic familiarity with concept maps. Submissions were made as pdf files. All concept maps, along with the accompanying descriptions, were downloaded for data processing and analysis. The research study was approved by the Institutional Review Board of the university.

### B. Data processing

All student identifiers were anonymized, and the submitted PDF files were also double-checked to ensure complete anonymization. The data were organized in an Excel sheet, and the components of each concept map and the accompanying description were recorded in two separate columns. Of the 86 students enrolled in the course, 15 did not submit the assignment, resulting in 71 concept maps available for initial processing. Two researchers independently conducted the holistic scoring of the concept maps proposed by Mary Besterfield-Sacre et al. (2004) and categorized the maps into three levels of comprehensiveness.

The maps were evaluated based on three criteria - comprehensiveness, organization, and correctness as defined in table I. For each criterion, maps were scored holistically on a scale from 1 to 3. Pluses and minuses were used to further differentiate the quality of maps (e.g., a good map might receive a 2+ or 3−, whereas an outstanding map might receive a 3 or 3+). In addition to assigning scores, both researchers provided critical comments for each map, noting its strengths and weaknesses. The scores were then mapped onto a nine-point scale (1 being the lowest and 9 the highest), and the average score for each map was calculated. Based on these averages, the concept maps were categorized into four levels to indicate comprehensiveness as a whole: Level 1 (1–3), Level 2 (3.1–5), Level 3 (5.1–7), and Level 4 (7.1–9). Disagreements were identified in case of level identification among the two researchers and there was a total of 29 disagreements. These cases were revisited by examining the criteria-based scores and comments, and disagreements were resolved through discussion of each criterion for the respective concept maps.

Concept maps that were difficult to read and follow, lacked meaningful concepts or connections, or did not align with the context of the assignment were classified as Level 1. In total, eight concept maps fell into this category. Two of these eight concept maps were retained as exceptional cases such as - although they were not structurally organized as concept maps with multiple interconnected concepts, they contained significant examples or explanations that can be helpful for the knowledge categories. Therefore, six concept maps were excluded, and the final dataset consisted of 64 concept maps for analysis.

TABLE I. Concept map scoring criteria

| Criteria | Description |
|---|---|
| Compre-hensiveness | Comprehensiveness referred to the student's ability, as represented in the map, to define the subject area and demonstrate the breadth and depth of knowledge. Specifically, the researchers assessed whether the map covered multiple aspects of generative AI, such as applications, limitations, ethical concerns, underlying technologies, and examples or explanations. |
| Organi-zation | Organization reflected the student's ability to systematically arrange concepts, including the hierarchy of concept placement and the integration of branches. The researchers assessed whether the connections and overall flow of the map were making sense and whether meaningful branches and cross-links were present. |
| Correct-ness | Correctness captured the accuracy of the information presented on the map, including the level of conceptual sophistication, the use of appropriate terminology, and the presence of misconceptions. The researchers assessed whether the concepts were accurate within the context of the assignment and whether the content was not too generalized or disconnected from the educational context. |

### C. Data analysis

*a) Core components identifications and hierarchical component clustering*

To explore how concept maps collectively represent students' mental models of GenAI (RQ1), two analyses were conducted. First, the core components of the concept maps were identified and their frequencies calculated across all 64 concept maps. Two researchers created a codebook in inductive way to code for the presence of components in the concept maps. The codebook included 16 codes in total. This analysis shows concepts students associated with GenAI, providing a bird's eye view of which aspects of GenAI are most salient in students' minds and which are absent or underrepresented (Fig. 1).

Second, hierarchical clustering was conducted to identify recurring patterns in how students grouped and connected concepts (fig. 2). This method identified what students know about GenAI and how they organize that knowledge. The resulting dendrogram visualizes the clusters as a tree structure, where each leaf represents an individual component and the branches illustrate how components merge into broader clusters. The height at which two branches merge represented on the vertical axis as distance indicates the degree of dissimilarity between them - concepts or clusters that merge at low heights are closely associated in the concept maps, while those merging at greater heights are more loosely associated. Each cluster representing different characteristics of students' mental model discussed in the Findings section.

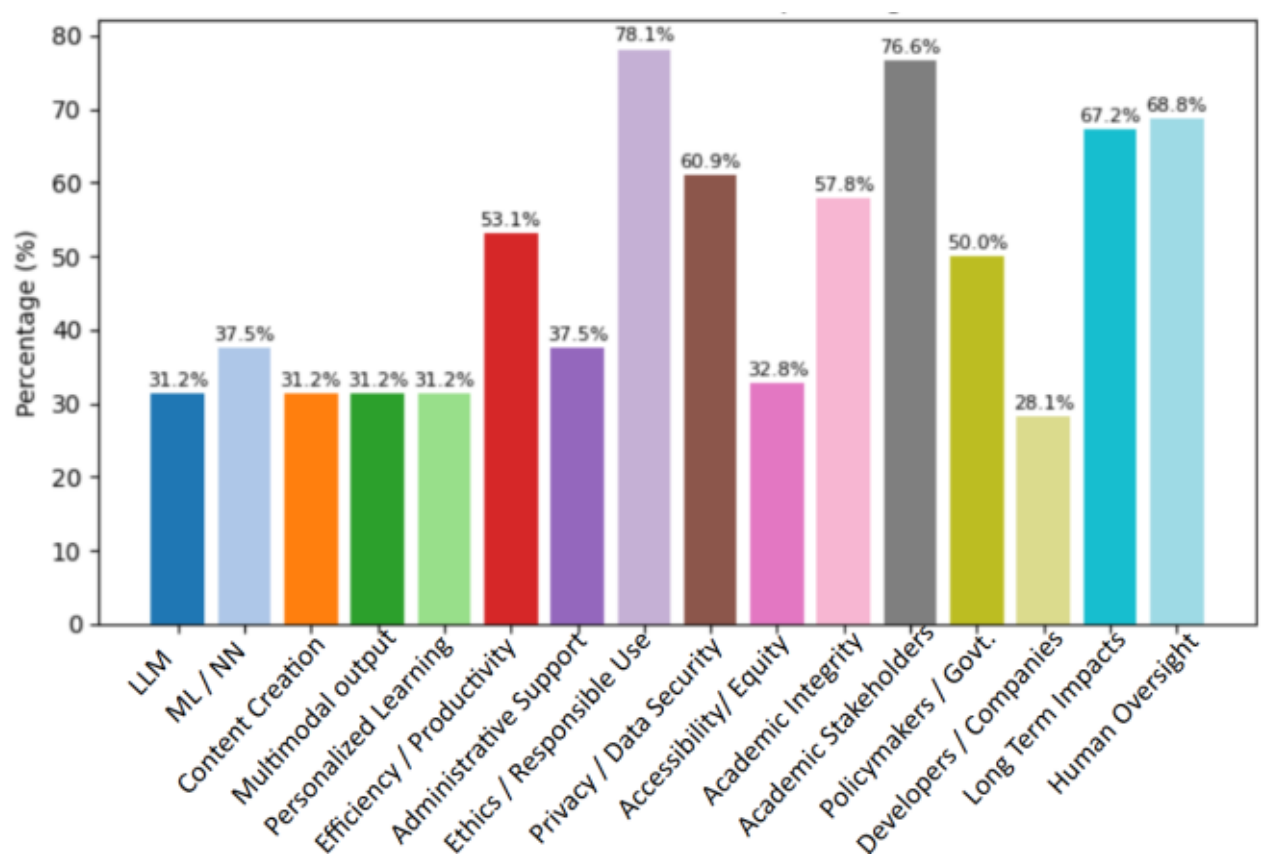

Fig. 1. Frequency analysis of components in GenAI concept maps.

### b) *Qualitative analysis of knowledge type*

To investigate what knowledge types shaped the mental models (RQ2), two researchers qualitatively analyzed each of the concept's maps and the description. For each knowledge type, set of criteria as questions were defined and discussed as shown in Table II.

TABLE II. KNOWLEDGE TYPES AND GUIDING QUESTIONS

| **Knowledge type** | **Guiding questions** |
|---|---|
| Declarative knowledge (knowledge of "WHAT?") | (i) Is the student's understanding limited to facts only with no process or judgment?<br>(ii) Does the student use only definitional language like "is a", "is called", "is defined as", "includes", "such as"?<br>(iii) Does the map show concepts without any explanation of how they relate to each other? |
| Procedural knowledge (knowledge of "HOW?") | (i) Are there processes, sequences (e.g. 3 step hierarchy of concepts indicating depth)?<br>(ii) Does the map include concepts showing how things work?<br>(iii) Does the edges have labels directing (leads to, produces) cause and effect relationship? |
| Conditional knowledge (knowledge of "WHEN and WHY?") | (i) Does the student make a judgment and sound thoughtful?<br>(ii) Does the description or the concept map include words like "should/ shouldn't", "must/mustn't", "need to", "have to"?<br>(iii) Does the concept map shows consequences/ concerns on a specific situation? |

## III. Findings

### A. *Mental model characteristics*

The hierarchical cluster analysis of student concept maps revealed five distinct characteristics of student's mental model answering RQ1. The clustering patterns suggest that students' mental models vary in aspects such as, some organize knowledge around how GenAI works, others around what it can do for them, and others around the broader responsibilities it demands from individuals, institutions, and societies. Based on these observed patterns, two researchers assigned descriptive labels to the five mental model categories. Future qualitative analysis of the concept maps is needed to better understand the nature of these connections. The five characteristics of student's mental model are discussed below:

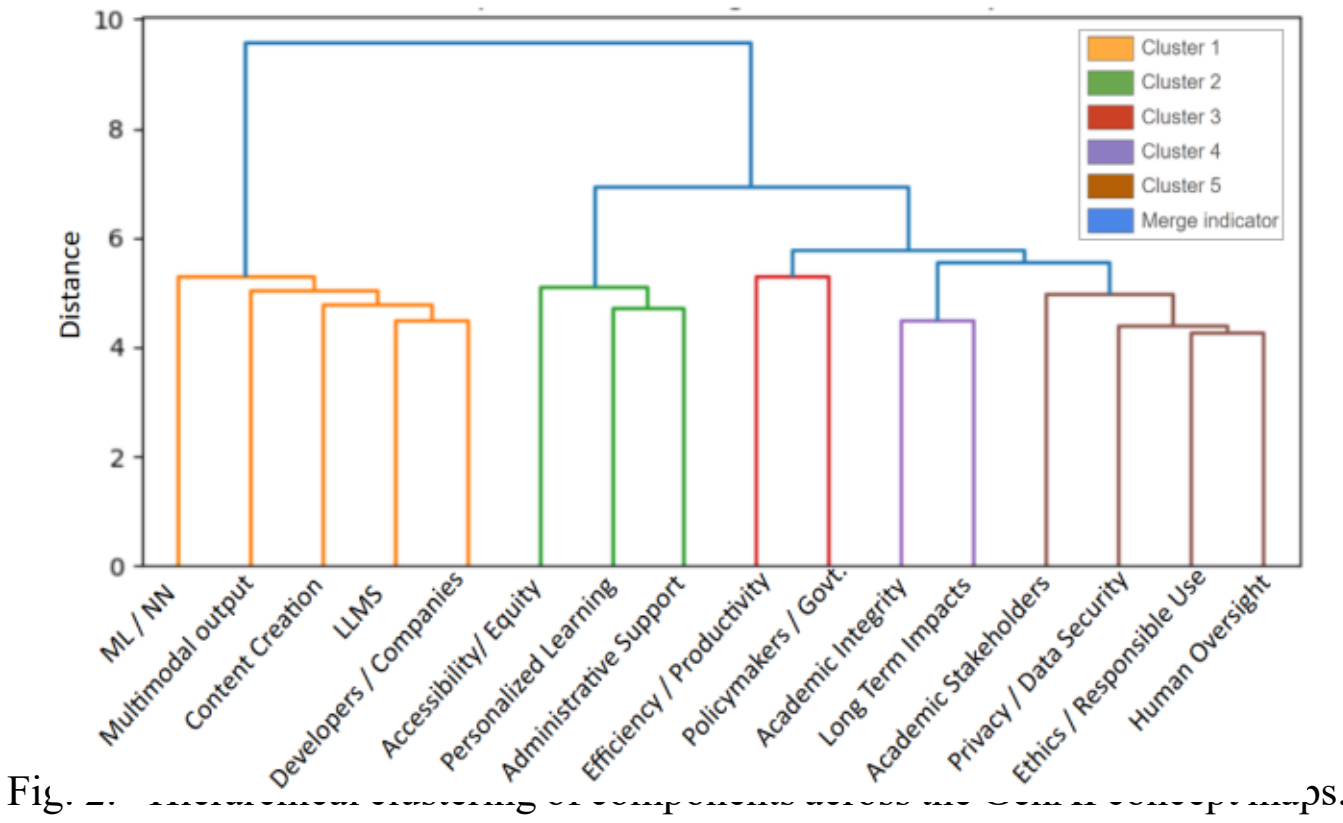

Fig. 2. Hierarchical clustering of components across the GenAI concept maps.

*GenAI through technical architecture:* Cluster 1 shows the concept maps grouped machine learning, neural networks, and large language models with its abilities such as content creation and multimodal output. The cluster also included Developers/Companies. These patterns suggest that students often connected GenAI's underlying technologies with its development context. This reflects a process-based mental model where technical architecture and generative capability are seen as inseparable.

*Focus on educational and accessibility aspects:* Cluster 2 indicates students group personalized learning, administrative support, and accessibility together, suggesting they perceive GenAI primarily as something that lowers barriers and supports learning in educational contexts. This cluster reflects a tool-based mental model where GenAI is seen as something that adapts to individual needs, reduces workload, and makes education more inclusive. Notably, this cluster is at a moderate distance from Cluster 1, indicating that students who think about GenAI's educational benefits do not necessarily connect those benefits to its technical foundations.

*Transition from productivity to regulatory awareness:* Cluster 3 demonstrates, rather than treating productivity and efficiency as a purely personal benefit, students included policymakers and government together with it in their concept maps. This pattern suggests a transitional mental model - students are beginning to move beyond individual tool use toward recognizing GenAI as a technology with consequences that require governance. However, the relatively small size of this cluster suggests this awareness remains limited among the student population.

*Consequence awareness:* Cluster 4 has academic integrity with long-term impacts, which makes sense as students who consider academic integrity such as plagiarism, cheating as a concern will also think about its long-term impacts. This pairing reflects a consequence aware mental model where students recognize that decisions made today about GenAI use specially in academic settings have implications that extend far into the future.

*Integrated mental model:* Cluster 5 shows students group ethics, privacy, human oversight, and academic stakeholders together, reflecting an understanding that responsible GenAI use is not an individual concern but a collective concern involving multiple stakeholders. The presence of human oversight alongside ethics and privacy suggests students have solution-oriented thinking to these concerns. Cluster 5 merges with cluster 4 indicating an integrated and risk-aware mental model where consequence awareness and ethical governance are two sides of the same coin in how students conceptualize the risks and responsibilities of GenAI.

## B. *Knowledge types shaping mental models*

The goal is to examine the types of knowledge reflected in the concept maps as mentioned in RQ2. All maps include components indicating declarative knowledge, while 25 concept maps show procedural knowledge and 17 reflect conditional knowledge. However, these counts in isolation provide limited insight. Understanding how this knowledge types co-occur offers a more meaningful interpretation. To address this, Figure 3 presents a Venn diagram illustrating the overlaps among the three knowledge types. Additionally, table III provides representative evidence from the concept maps corresponding to declarative, procedural, and conditional knowledge.

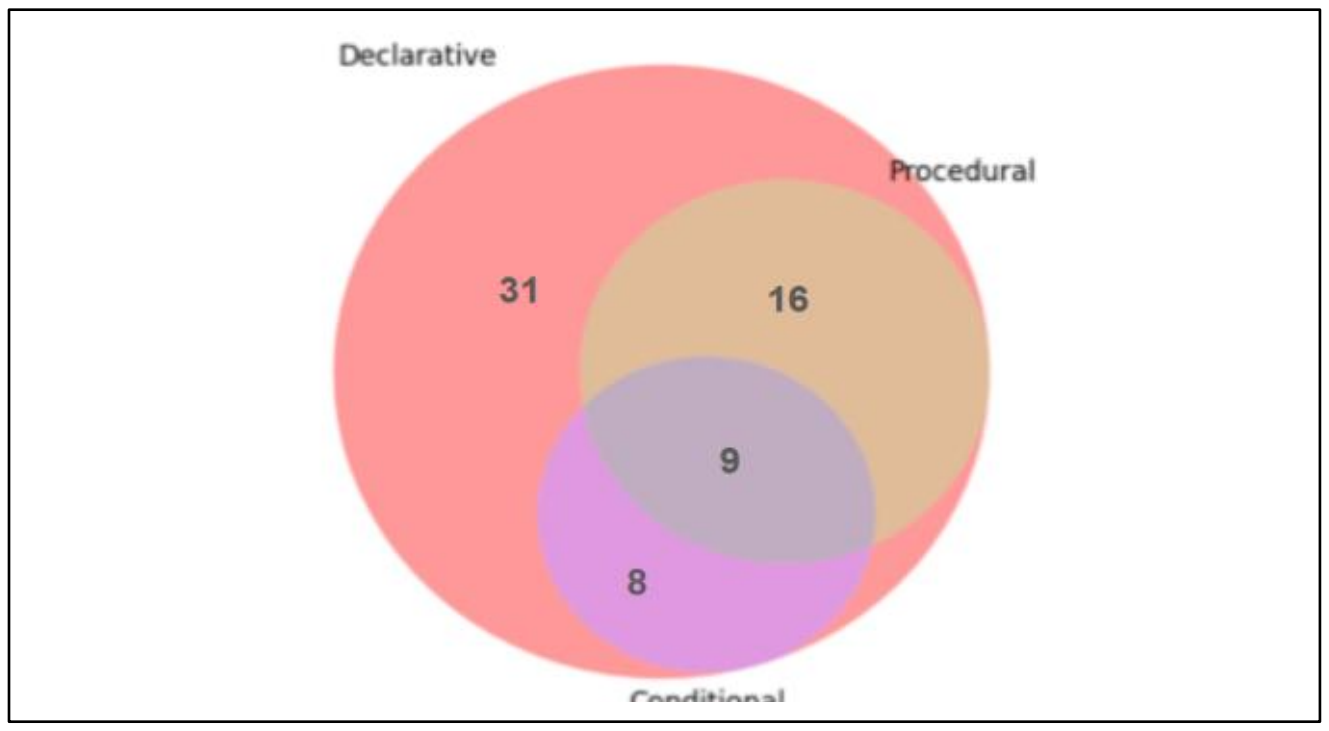

Fig. 3. Overlap of knowledge types.

*Integrated knowledge:* Only 9 of the students' concept maps were shaped by all the three types of knowledge. Indicating the most sophisticated group. Their mental models are integrated and multilayered.

*Declarative knowledge:* 31 concept maps were declarative only. These students only listed and categorized concepts pure factual/definitional knowledge with no process reasoning and no evaluative judgment leading to the most surface-level mental models.

*Declarative and conditional knowledge:* 8 concept maps representing combination of declarative and conditional knowledge with no procedural knowledge, these students know what GenAI is and have judgments regarding it regarding when and why but show no process understanding such as how a concept leads to another concept and how GenAI works. They can name concepts and express concerns but can't trace how technology works.

*Declarative and procedural knowledge:* A total of 16 concept maps is a combination of declarative and procedural. These students understand the process of how GenAI works around different concepts and can name the components but show no evaluative or ethical judgment in their maps leading to technically oriented but not critically oriented mental models.

TABLE III. KNOWLEDGE TYPES AND EXAMPLE CONCEPTS COLLECTED FROM THE CONCEPT MAPS.

| Knowledge type | Example |
|---|---|
| Decla-rative | i. [GenAI] > [AI systems that can create new and original content]<br>ii. [Application] > [Summarization], [Advising], [Content generation]<br>iii. [Limitation] > [Academic integrity], [AI bias], [Plagiarism] |
| Proce-dural | i. [GenAI in Education] > [tutoring] > [summarization, comprehension]<br>ii. [GenAI is a program to think on its own, creates new material, trained through resources from various creators]<br>iii. [Training data] > [Data quality] > [Quality of data directly impacts quality of results]<br>iv. [It can adapt to student's particular needs] > by > [helping students writing assignments] |
| Condi-tional | i. [Stakeholders] > [Instructors] > [Instructors must ensure students don't cheat]<br>ii. [Challenge] > [The challenge isn't just adopting AI, it's making sure it's implemented responsibly, equitably, and with the right safeguards in place]<br>iii. [Concerns] > [Just like with any tool, you need to consider what you're using it for], [Many legal questions are unanswered regarding the intellectual property]<br>iv. [Solutions] > [Double check AI outputs], [Be open and clear about AI use] |

# IV. Discussion

The findings of this study shows that students do not hold a unified mental model of GenAI, but rather construct different concept maps with different levels of engagement with the technology. The large distance between the technical and social-regulatory clusters suggests a disconnect between understanding how GenAI works and recognizing its broader consequences, indicating that technical literacy and ethical awareness may be developing separately. Similarly, emergence of a transitional mental model in Cluster 3, where students begin to associate productivity and efficiency with the need for regulatory oversight goes beyond personal utility. The most developed mental model, reflected in Cluster 5, demonstrates that a subset of students has begun to think about GenAI through a collective and solution-oriented lens, connecting ethics, privacy, human oversight, and stakeholder responsibility into a coherent framework. Together, these patterns suggest that students are at very different stages of conceptual development regarding GenAI, and that education and GenAI guidelines has yet to bridge the gap between technical understanding and ethical reasoning. The dominance of declarative knowledge also carries an interpretation. When declarative knowledge appears without accompanying procedural or conditional dimensions, it signals a mental model that is wide but shallow. Moreover, the fact that declarative knowledge alone characterized a subset of students - those whose maps contained rich lists of concepts but little evidence of process reasoning or evaluative judgment points to a surface-level engagement with the material. This aligns findings from the knowledge structure literature suggesting that declarative knowledge is the most readily activated and most easily assessed form of knowledge [13, 14], particularly in novice learners encountering a domain for the first time. Findings from this study therefore informs the need for educational approaches and implications for different academic stakeholders who can bridge these conceptual domains.

### A. Implications for faculty and students

Students in Cluster 1 who understand GenAI through its technical architecture may benefit from having knowledge on human and societal contexts. Students in Cluster 2 who see GenAI primarily as an educational tool may need prompting to reflect on who has access to those tools, under what conditions, and with what risks. Faculty can use the cluster structure as a diagnostic framework such as identifying where students' mental models lie and designing learning activities that stretch their thinking across cluster boundaries. The goal is for students to develop what might be called an integrated mental model - one that holds technical, educational, ethical, and governance dimensions simultaneously, recognizing them as interdependent rather than separate concerns.

### B. Implications for researchers

The use of concept maps and hierarchical clustering as a methodology for studying mental models can be extended across different institutional contexts, disciplines, and student populations. Researchers might investigate whether the five mental model characteristics identified here are stable across cultures and educational systems, or whether they vary significantly by context. Longitudinal studies [15] tracking how students' mental models evolve over the course of a degree program or in response to specific interventions would also be valuable. Future research could explore the relationship between mental models and GenAI use behavior, examining whether students with more integrated mental models make more reflective and responsible choices in practice.

### C. Implications for design of guidelines for use

Current guidelines at many institutions tend to focus narrowly on academic integrity - addressing plagiarism, disclosure, and permitted use which aligns with only one dimension of how students conceptualize GenAI. The findings from this paper suggest that effective guidelines need to be broader and more layered, addressing not only what students are allowed to do with GenAI, but why responsible use matters at a technical, educational, ethical, and societal level. Guidelines should be structured to reflect the five mental model characteristics identified, progressively building from awareness of what GenAI is and how it works, through its educational affordances and productivity implications, to its long-term consequences and the collective responsibilities it entails. This scaffolded approach to guideline design would meet students where their current understanding is while actively working to expand it toward more integrated and consequence aware thinking.

## V. Conclusion

This WIP paper used concept map analysis to understand how undergraduate students conceptualize GenAI use. From the preliminary data analysis, we identified five distinct mental model characteristics - technical-process based, educational tool based, transitional, consequence awareness, and integrated mental model. In addition, the concept maps were analyzed to identify the types of knowledge represented in students' understanding. Only a small subset of students held all knowledge types, suggesting that integrated mental models remain the exception. As a WIP study, these results are preliminary. Future work will complement concept map analysis with interviews and think-aloud protocols to validate the mental model characteristics identified here.

## Acknowledgment

This work is partly supported by U.S. NSF Award# 2439459, 2439460, 2319137, 1954556, and USDA/NIFA Award# 2021-67021-35329. Any opinions, findings, and conclusions or recommendations expressed in this material are those of the authors and do not necessarily reflect the views of the funding agencies.